\documentclass[11pt]{article}
\usepackage{amsmath}
\usepackage{amssymb}
\usepackage[dvips]{graphicx}
\usepackage{color}
\usepackage{graphicx}
\usepackage{graphics}
\usepackage{epsfig}
\usepackage{hyperref}
\usepackage{fancyhdr}
\long\def\symbolfootnote[#1]#2{\begingroup%
\def\thefootnote{\fnsymbol{footnote}}\footnote[#1]{#2}\endgroup}
\begin{document}
%
%
{\Large\center Behavior of aqueous Tetrabutylammonium bromide - a combined approach of microscopic simulation and neutron scattering}  \\

\noindent Debsindhu Bhowmik$^{1}$ \\

\vspace{0.1cm}

\noindent $^1$ Computational Science and Engineering Division, Oak Ridge National Laboratory, USA \\

\noindent *email: bhowmikd@ornl.gov \\

%
%
\vspace{0.4cm}

\begin{abstract}
Aqueous solution of tetrabutylammonium bromide is studied by quasi-elastic neutron scattering, to give information on the dynamic modes involving the ions present. Using a careful combination of two techniques, time-of-flight (TOF) and neutron spin echo (NSE), we decouple the dynamic information in both the coherently and incoherently scattered signal from this system. We take advantage of the different intensity ratio of the two signals, as detected by each of the techniques, to achieve this decoupling. By using heavy water as the solvent, the tetrabutylammonium cation is the only hydrogen-containing species in the system and gives rise to a significant incoherent scattered intensity. The dynamic analysis of the incoherent signal (measured by TOF) leads to a translational diffusion coefficient of the cation as that is in good agreement with previous NMR, neutron scattering and tracer diffusion measurements. The dynamic analysis of the coherent signal observed at wave-vectors $<$ 0.6\AA$^{-1}$ (measured by NSE) leads to a significantly slower diffusive mode, by a factor of 2. This study explains that apparent difference and the reasons for that. 
\end{abstract}

\newpage

\section{\label{intro}Introduction}
%
%
The natural abundance and numerous application of the hydrophobic ions in the domain of biology, chemistry or technology, makes it an active field of study for decades. One of the most investigated systems of this kind are symmetric tetraalkylammonium (TAA) (C$_{n}$H$_{2n+1})_{4}$N$^{+}$ ions~\cite{Bhowmik2012}~\cite{Bhowmik2011a}~\cite{Bhowmik2014b}~\cite{Malikova2012}. TAA halides are considered to be model system to study the behavior of hydrophobic ions in solution. These salts allow to go beyond the 1:1 alkali or inorganic electrolytes to hydrophobic systems and provides a model system to study the combined effect of the hydrophobic effect on the apolar surface (short range van der Waals force) and long range electrostatic force. Changing the hydrophobicity by modifying the hydrocarbon chain length makes them even more interesting. This hydrophobicity is also assumed to affect the solvent structure and its dynamics~\cite{Somsen93}~\cite{Turner95_102}~\cite{Novikov01_91}. Taking into account the concept of structure making/ breaking~\cite{Frank57} one can differentiate the solvent molecule in two parts. One are those solvent molecules which are directly attached to the ions and rest are the molecules surrounding the ions. Our primary objective is to study the solute dynamics which can be affected by neighboring solvent molecules. To reduce bulk solvent effect we have used a very high concentration where only a single hydration layer can be formed around the solute. But it is always difficult to estimate the definite number of water molecules to form the first independent hydration. From OH and CH Raman spectra analysis~\cite{Green87} of tetrabutylammonium (TBA) cation, it can be shown that the hydrogen bond defect probability of first hydration layer starts around a concentration corresponding 56 water molecules per cation~\cite{Turner94_101}. Keeping that in mind we have studied aqueous TBABr solution  by molecular dynamics (MD) simulation with a vast concentration range (0.05M to 2M) that also covers 1m (ion:water=1:56). Here we only present the 1m result for aqueous TBABr solution and at the same time we make a comparison with our quasi-elastic neutron scattering (QENS) result. It is interesting to note that earlier hypernetted chain (HNC) integral equation is used to explain the small angle scattering data~\cite{Calmettes92} with a solvent averaged spherically symmetrical potentials for TBA ions which is shown to work only greater than $\sim$20\AA$ $ and does not speak about the atomic or molecular dynamics. The main conclusions are cation hydrocarbon chains are fully stretched, no hydrophobic bonding from cation-cation correlation and water molecules, anion or other alkyl chain can penetrate into cation. Our result go beyond the large spatial scale to very small intra-atomic distance reproducing well the experimental spectra deduced by different techniques and to the best of our knowledge, this is the first detailed dynamical study (in comparison with quasi-elastic neutron data) where TBA$^{+}$ is considered as completely non-rigid. On the context of ion pair formation and penetration, it is proposed that different kind of association of TAA occurs with bromide or iodide (water structure-enforced ion pairing) and chloride~\cite{Diamond63}. By an all atom explicit model [from TMA (tetramethylammonium bromide) to TPA(tetrapropylammonium bromide)], it is shown that that this ion pairing follows inverse Hofmeister series~\cite{Heyda10} which is also supported by the study of mutual diffusion~\cite{Kim73}. By dielectric experiment, it is suggested that the bromide ion can penetrate upto the distance same as TMA~\cite{Buchner02}. In our study we will also try answer these aspects in view of TBA. For dynamics, many previous studies suggest that TBA cations diffuse following continuous diffusion law with no orientational motion. Although looking at the size of TAA cations it is not so obvious. The present article will deal with all these solute and also solvent behavior in aqueous medium at 1m concentration.
It is structured as follows - after a brief introduction, in section 2, simulation and experimental techniques are described. Then result is discussed in two parts (Structure and Dynamics) where each segment is divided into a) solute and b) solvent. At the end we sum up to conclude this work.

\section{\label{tech}Technique}
%
%

\subsection{\label{tech_sim}Simulation}
%
%
Molecular dynamics (MD) simulation (using the code DL POLY 2.18~\cite{Smith07}) are performed on aqueous solution of NaBr, TMABr and TBABr to study the solute and solvent dynamics~\cite{Bhowmik2014b}~\cite{Bhowmik2011a}~\cite{Bhowmik2012}. The three different aqueous solutions are compared to see how the solvation structure is changed as we move from small simple salts (ex: NaBr) to bigger hydrophobic TBABr via TMABr. Considering TBA cation, all the atoms in the ion are treated individually with maximum number of degrees of freedom by incorporating all chemical bonds (harmonic), valence angles (harmonic) and dihedral angles (cosine) potentials in our simulation. A non-polarizable force field potential is used for TBA$^{+}$. Except the charges, all the other parameters are taken from Generalized Amber Force Field (GAFF)~\cite{Amber_10}. The atom charges are determined by Hartree-Fock method (for nonpolarizable force field). Then incorporating all the parameters, these charges are modified (minor) by Antechamber (AMBER routine)~\cite{Heyda10}. The sodium, bromide parameters are taken from literature~\cite{Koneshan98_102}~\cite{Horinek09}~\cite{Lee96}~\cite{Joung08}~\cite{Markovich96_7} and rigid SPC/E model is used for solvent water~\cite{Berendsen87_91}. This essentially leads to two kinds of intermolecular pair interaction potentials (non-bonded) in the system which are long range electrostatic coulomb and short range 6-12 Lenard-Jones (L-J) and thus each atom in the system experiences a resultant potential which can be written as 
\begin{eqnarray}
\label{total_potential_1} V_{ij} = \frac{q_{i}q_{j}}{4\epsilon_{0}r_{ij}} + 4\epsilon_{ij}[(\frac{\sigma_{ij}}{r_{ij}})^{12} - (\frac{\sigma_{ij}}{r_{ij}})^{6}]
\end{eqnarray}
where $\sigma_{i}$, $\epsilon_{i}$ are L-J parameters and charges are represented by q$_{i}$. Lorentz-Berthelot rule is applied for calculating the pair parameters (energy and size) between unlike atoms (for i$\neq$j) i.e. $\epsilon_{ij}$=$(\epsilon_{i}\epsilon_{j})^{\frac{1}{2}}$ and $\sigma_{ij}$=$\frac{(\sigma_{i}+\sigma_{j})}{2}$. 

First the initial simulation box is constructed with equidistantly placed ions (cations and anions) and then the water molecules are incorporated with random orientation using a prior idea of the volume of ions and water molecules. Three dimensional periodic boundary condition is employed with a cutoff radius equal to half the box-size for real space of electrostatic potentials (also same distance for non-bonded interactions) using 3D ewald sum for long range coulombic interactions. The entire system is then allowed to equilibrate for 600ps in the NPT ensemble. It is run again in NPT and NVT ensemble respectively and the final simulation is carried out in NVE ensemble. The Nos\'{e}-Hoover thermostat [with relaxation constant 0.5ps (in NVT)], temperature = 298K and pressure = 1atm couplings are employed and for rigid water molecules all bonds were constrained by SHAKE algorithm. The total simulation time is 3.4ns in NVE with 1fs timestep. The atom trajectories are saved every 0.1ps producing 34$\times$10$^{6}$ frames in total. The MD simulated trajectories (atom position as a function of time) are then analyzed using nMoldyn~\cite{nMOLDYN}~\cite{Bhowmik2014b}~\cite{Bhowmik2011a}.

Different concentrations are compared with experimental results (from 0.05M to 2M) to check whether the simulated system density is reproduced well (difference is $<$0.2\%). In this article we will present our MD simulation result for one concentration (1 molal or 8 cations and 8 anions in 448 water molecules). At the same time, we will also compare with our quasi-elastic neutron scattering (QENS) experimental data to see how far this model can be considered to be realistic (for TBABr with density 1.016 gm/cm$^{3}$). 

\subsection{\label{tech_exp}Experimental}
%
%
As stated earlier our experimental data comes from neutron scattering where it is a common practice to measure the spectra as function of Q. The neutron scattering technique is a valuable tool to explore molecular structures and dynamics of biomacromolecules~\cite{Bhowmik2015a}~\cite{Bhowmik2016}~\cite{Bhowmik2015b}~\cite{Bhowmik2015c}~\cite{Dhindsa2015}~\cite{Dhindsa2016}~\cite{Perera2016}~\cite{Shrestha2016}~\cite{Shrestha2015}~\cite{Shrestha2015a}~\cite{Shrestha2015b}, polymers~\cite{Arbe2016}~\cite{Bhowmik2014}~\cite{Bhowmik2014a}~\cite{Bhowmik2015} or polyelectrolytes~\cite{Bhowmik2012}~\cite{Bhowmik2014b}~\cite{Bhowmik2011a}~\cite{Malikova2012} in different length and time scale.  Here Q presents the reciprocal wave vector (inverse of distance) which is the difference in momentum transfer~\cite{Bee}~\cite{Squires} (this is the reason for invoking the concept of Q in the analysis of our simulation). The small angle neutron scattering (SANS), covering Q range between 0.037\AA$^{-1}$ to 0.302\AA$^{-1}$, is performed on PAXE and neutron diffraction (ND) technique is carried out on G4.1 spectrometer in LLB-Orphee, Saclay, France. For the dynamics, We make use of both neutron spin echo (NSE)~\cite{Mezei} and time of flight (TOF)~\cite{Bee}. While by NSE, we measure the intermediate scattering function I(Q,t) in time domain; the TOF records the dynamical structure factor S(Q, $\omega$) which is the fourier transform of I(Q,t) in energy domain. NSE experiments are carried out (up to a correlation time of 1100 ps from 0.2\AA$^{-1}$ to 1.6\AA$^{-1}$) on RESEDA and MUSES spectrometers on FRM-II, Munich, Germany and LLB-Orphee, Saclay, France respectively while TOF measurements (from 0.49\AA$^{-1}$ to 1.97\AA$^{-1}$ with a resolution of HWHM = 50 $\mu$eV) are performed on MIBEMOL in LLB-Orphee, Saclay, France. For further technical detail one can consult our previous work~\cite{Bhowmik2012}.

\section{\label{results}Results and discussion}
%
%
In this section, we present our MD results mainly for 1m aqueous TBABr solution to study the statics and dynamics of solute and solvent. At the same time, we compare with our neutron data to verify how realistically our model can reproduce experimental results. While explaining the results we will not show extensively the experimental data analysis procedure (as it is beyond the limit of this article and also discussed in our other article~\cite{Bhowmik2012}), instead we will show the final results from the experiments which are comparable to the simulation and emphasize more on the latter. In each of the following sections of structure and dynamics, we begin by studying the solute and then the solvent.

\subsection{\label{structure}Structure}
%
%

\subsubsection{\label{structure_solute}Solute}
\begin{figure}[htbp]
\begin{center}
\includegraphics[width=0.55\textwidth,angle=-90] {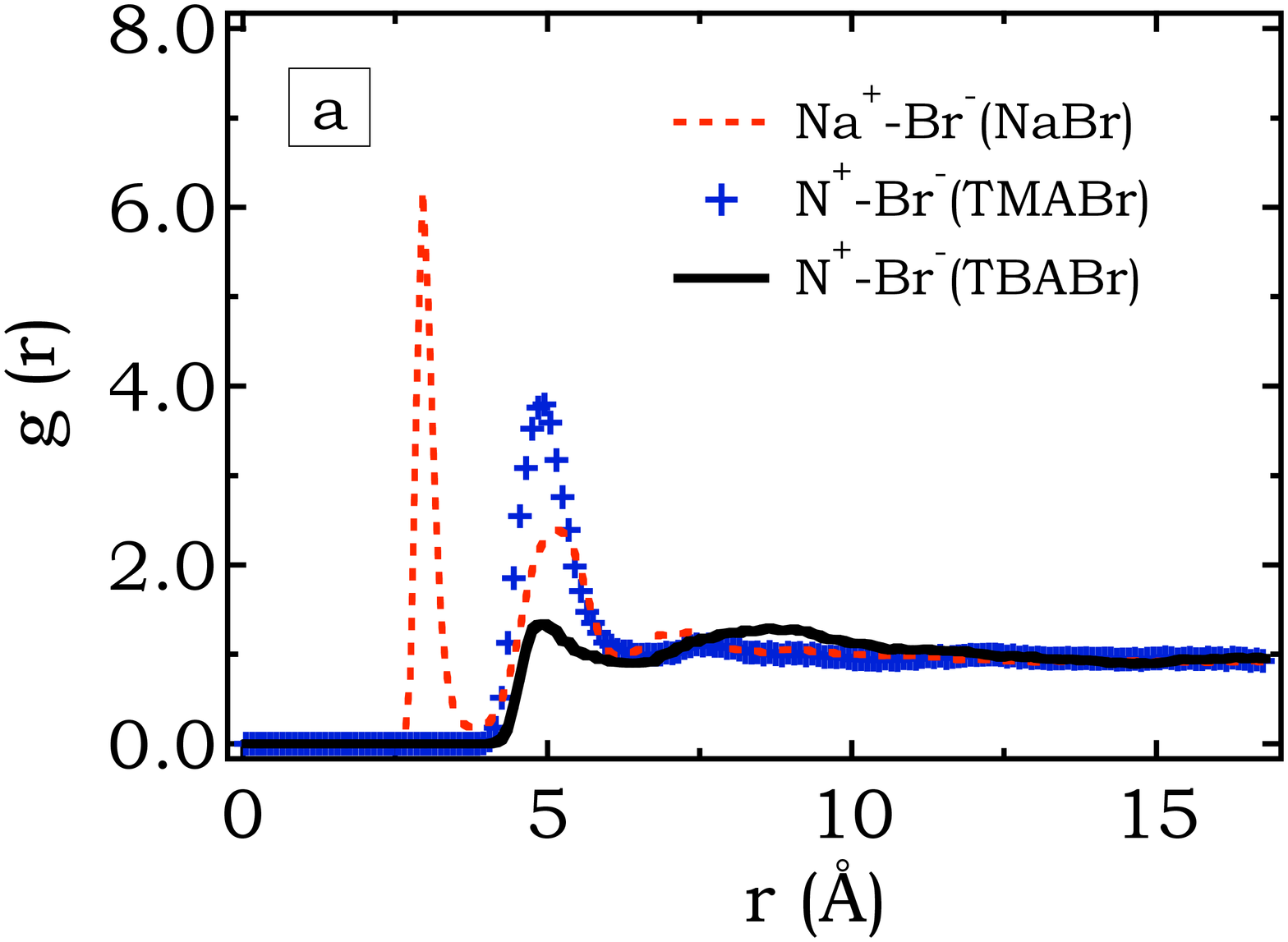}
\includegraphics[width=0.65\textwidth,angle=-90] {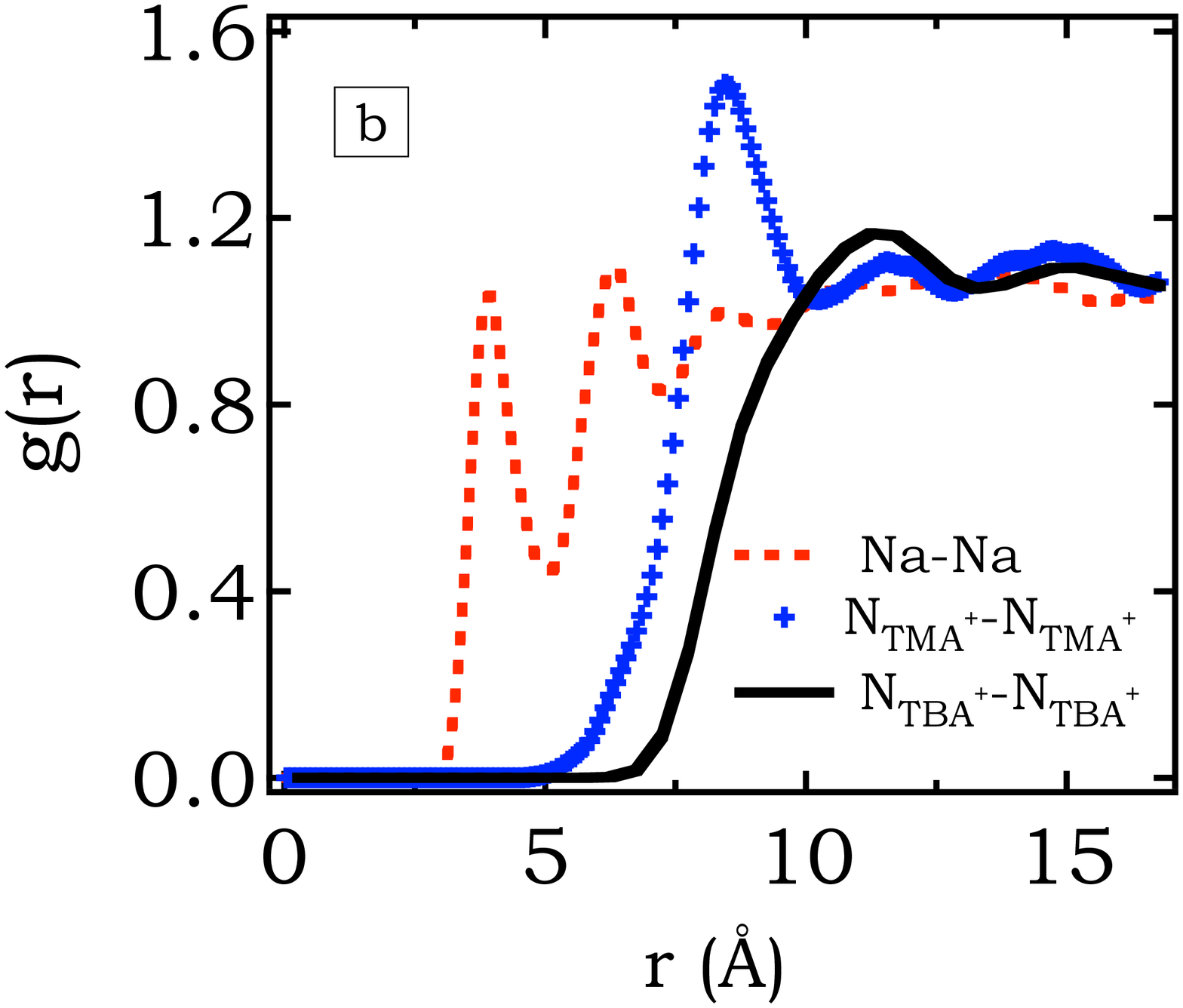}
\vspace{0.1cm}
\caption{ion-ion radial distribution function of 1m (ion:water = 1:56) TBABr, TMABr and NaBr solution. (a) how bromide anion is associated to different cations and penetration in case of TBA (b) Some interaction is present around 11\AA$ $ for TBA cation which is largely nullified by presence of oppositely charged anions.}
\label{gNBr-1mTBABr}
\label{g_r-S_Q}
\end{center}
\end{figure}

%
%
Looking into Figure \ref{gNBr-1mTBABr}(a), we have the cation-anion pair radial distribution function (RDF), where bromide ions have a distinct correlation peak with Na$^{+}$ (of NaBr) or N$^{+}$ of TMABr but in case of TBABr, significantly less pronounced peak is observed. It also suggests that as the correlation between N$_{TBA^{+}}$ and Br$^{-}$ goes upto $\sim$5\AA, there is probability that bromide can penetrate into cationic hydrocarbon chain of TBABr. Moving to the cation-cation pair-distribution function (Figure \ref{g_r-S_Q}(b)), we see a small correlation exists around 11\AA$ $ for TBABr compared to TMABr. The different cation-cation peak for TMA and TBA cation can be explained as follows. The origin of this peak is due to the cationic repulsion. For TBA$^{+}$, a rough estimate of the distance between two consecutive nitrogen atom (cation CoM) is $\sim$13\AA. Now the last hydrogen of TBA$^{+}$ is around 6\AA$ $ which indicates that two neighboring cations are very closely placed. At the same time recalling the fact that the anion can penetrate up to $\sim$5\AA$ $, it indicates that coulombic repulsion between the cations are largely neutralized by the presence of anions (supporting cation-anion pair formation) and leaves a resultant weak repulsive interaction. The fact is further supported by our SANS conclusion~\cite{Bhowmik2012} and also from the calculation of structure factor, discussed later. This is not the case for TMA$^{+}$. Although the bromide ions are also present there but they neither go into the cation hydrocarbon chain nor form ion-pair. And thus coulombic repulsion due to cations is not reduced like TBA$^{+}$. For simple ion (as Na$^{+}$ presented here), the situation is rather simple. The two cations can not come closer than $\sim$4\AA$ $ and their intermediate space is completely filled up by the anions. The dip in Na $^{+}$ RDF is due to the Br$^{-}$ ions (i.e. peak in Na$^{+}$ and Br$^{-}$ RDF is at the dip of Na$^{+}$ and Na$^{+}$ RDF). 

\begin{figure}[htbp]
\begin{center}
\includegraphics[width=0.9\textwidth,angle=-90] {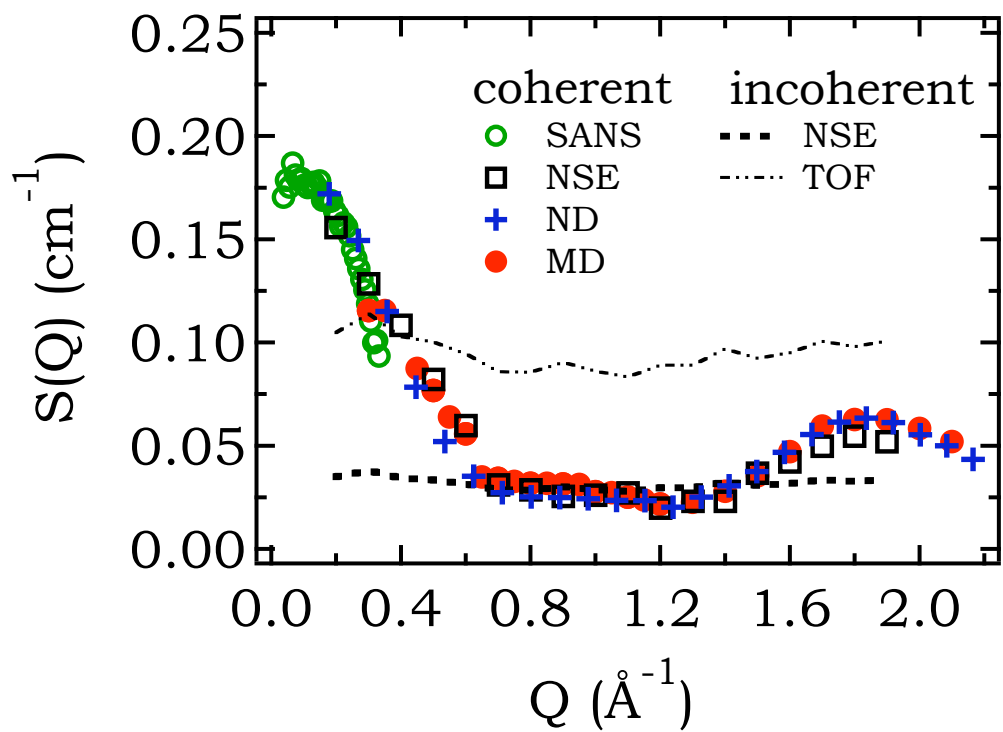}
\vspace{0.1cm}
\caption{static structure factor (cm$^{-1}$) of 1m (ion:water=1:56) TBABr solutions by different techniques. Note that the actual incoherent contribution (as shown by TOF) is reduced by a factor 3 for NSE measurement due to spin-flip. The agreement of our MD simulation result with experiments covers a large Q range}
\label{S_Q}
\end{center}
\end{figure}

In this last segment of this structural characterization, we verify whether the simulation field parameters are good enough to reproduce the static structure factor S(Q) as obtained from different experiments. We calculate the S(Q) by considering the positional correlation of all the atom at time t=0. In Figure \ref{S_Q}, we show the static structure factor S(Q) estimated by the MD-simulation and also by our other experiments. In Figure \ref{S_Q}, the static structure factor S(Q) (which is coherent in nature~\cite{Mezei}~\cite{Bee} in nature) is superimposed on incoherent~\cite{Mezei}~\cite{Bee} intensity coming from the system. Note that for NSE experiment, the true incoherent contribution is diminished by a factor of 3 because of the inherent neutron spin flip during incoherent scattering~\cite{Squires}. We show that MD simulated S(Q) is in accord with our other experimental measurements (such as neutron diffraction, SANS and NSE), covering a large Q range from Q=0.03\AA$^{-1}$ to 2.2\AA$^{-1}$. The increase in intensity of S(Q) at low Q (upto 0.65\AA) is mainly due to the cation-cation (self + different) correlation and at high Q ($\sim$1.9\AA$^{-1}$) is due to the solvent molecule correlation. It is also possible to calculate the S(Q), using the RDFs between all the atoms~\cite{Kunz92_96} ~\cite{Kunz91_95}. If only the RDFs of particle CoM (centre of mass) are taken (here it is central nitrogen atom for a fully stretched symmetric TBA$^{+}$), one can calculate the structure factor $S_{CoM}(Q)$. The difference between S(Q) and $S_{CoM}(Q)$ is that while $S_{CoM}(Q)$ determines the positional correlation between particle centre of mass, S(Q) is the product of form factor and $S_{CoM}(Q)$. Calculated structure factor ($S_{CoM}(Q)$) shows constant value and a small peak around Q=0.65\AA$ $ and then steadily decreases towards lower Q indicating the presence of some interaction (but not significant) which is also predicted from our earlier SANS experiment~\cite{Bhowmik2012}.

\subsubsection{\label{structure_solvent}Solvent}
\begin{figure}[htbp]
\begin{center}
\includegraphics[width=0.9\textwidth,angle=-90] {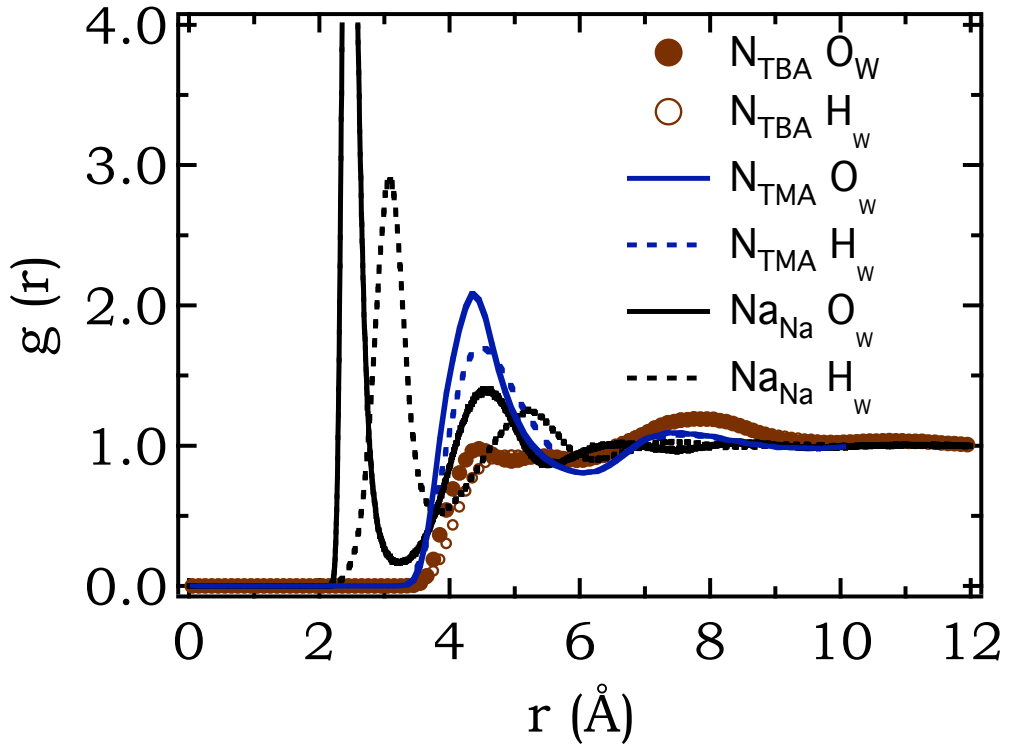}
\vspace{0.1cm}
\caption{cation-water Radial Distribution Function at 1m (ion:water = 1:56) concentration for TBABr, TMABr and NaBr solutions to show how the solvent water is structured differently as we move from simple (NaBr) to hydrophobic salts (TMABr, TBABr)}
\label{N-O_TAABr}
\end{center}
\end{figure}

%
%
A comparison to simple salts (NaBr) or smallest TAA cation (TMABr) is useful to see how differently the solvent structure is affected in case of TBABr. Here all the results are with a concentration of 1m (ion:water = 1:56). The most straight forward way to study the structure is to look into the RDF. In Figure \ref{N-O_TAABr}, we present the RDF between cation (N$^{+}$ of TBABr and TMABr or Na$^{+}$ of NaBr) and oxygen/ hydrogens of water. From Figure \ref{N-O_TAABr}, we can mainly draw two conclusions - i) the structures of water-shell around cation and ii) how much the water molecules can penetrate inside the TAA ions. In case of simple salts the hydration shell is distinct, well structured and the orientation of oxygen and hydrogens of water molecules are as expected i.e. oxygen atoms are closer than hydrogens with more probability because of the positive cationic nature of Na$^{+}$. This situation is not the same for hydrophobic aqueous TMA$^{+}$ ions. Here the oxygen and hydrogen atoms are at same distance from cation with almost equal probability (and the RDF peak is less intense than simple salts). This indicates that the water orientation is tangential. This conclusion is in agreement with earlier diffraction measurement~\cite{Soper92_77}~\cite{Turner95_102}. For TBA$^{+}$, the hydration shell is shifted even more ($\sim$7.8\AA) (as expected due to a larger size) in comparison to TMA$^{+}$($\sim$4.4\AA$ $ which is same as the second hydration sphere for simple salts) and RDF peak-height is also diminished. The The same figure also tells that for TBA$^{+}$, the water can penetrate up to the same distance ($\sim$3.4\AA) as TMA$^{+}$~\cite{Buchner02} though the first prominent hydration shell is farther away ($\sim$7.8\AA). Before this distance the water orientation has no preferential direction (this can be inferred from the fact that the g(N$_{TBA}$O$_{W}$) or g(N$_{TBA}$H$_{W}$) is non-zero constant value $<\sim$7.8\AA). This is because of several effects like electrostatic interaction of atoms in cation and the movement of the cation hydrophobic chains which influence the water molecules inside the four hydrocarbon arms. In case of TBA$^{+}$, water molecules can penetrate upto the third carbon of each arm. The last hydrogen of the TBA$^{+}$ is at $\sim$6\AA$ $ from the centre of mass (CoM). Using this RDF and integrating up to appropriate distance, it is estimated that up to the last hydrogen of respective TAA cation, there are $\sim$16 water molecules for TBA$^{+}$ and none for TMA$^{+}$ (size of TMA$^{+}$ is $\sim$3.4\AA). Note that the first hydration shell of TMA$^{+}$ and TBA$^{+}$ consists of $\sim$25 and $\sim$5O H$_{2}$O molecules respectively. Thus we find that as the cationic radius increase (from simple salts to TBA$^{+}$), the cation-water interaction becomes weaker as the peak of cation-water RDF diminishes. The presence of hydration shell around TAA ions suggests that the cations can stay inside a cage but as the solute size increases the solvation cage becomes weaker (as the RDF peak intensity of cation-water decreases from TMA to TBA)~\cite{Bhowmik2014b}~\cite{Bhowmik2011a}.  

\subsection{\label{dyn}Dynamics}
%
%

\subsubsection{\label{dyn_solute}Solute}
%
%

\begin{figure}[htbp]
\begin{center}
\includegraphics[width=0.9\textwidth,angle=-90] {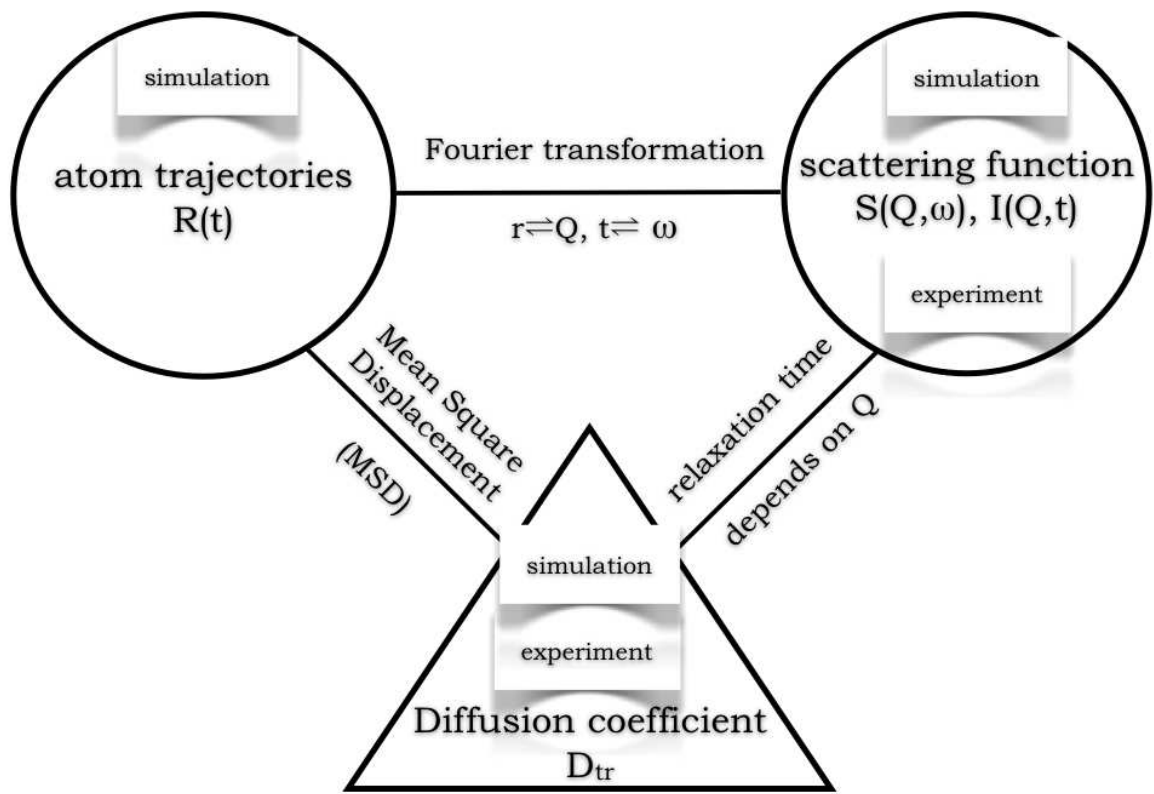}
\vspace{0.1cm}
\caption{A schematic diagram to show how we arrive at $D_{tr}$ by using different approaches both in simulation and in expriment}
\label{connection}
\end{center}
\end{figure}

\
\\
\bf{A. Translational motion}\normalfont
%
%
\\
\\
One can measure the translational dynamics of any particle by two different ways - either by i) analyzing individual atom motion or by ii) considering the CoM motion~\cite{Bhowmik2011a}~\cite{Bhowmik2014b}. The first way of determining translational diffusion coefficient $D_{tr}$ is well exercised. In case of MD simulation, here the individual atom motion is accessed by two different techniques - MSD (Mean Square Displacement) and intermediate scattering function I(Q,t)~\cite{Mezei}~\cite{Bee}. In Figure \ref{connection} we have shown how the translational diffusion coefficient ($D_{tr}$) can be calculated by different experimental or simulation approaches. 
In MSD for isotropic system in 3D, $D_{tr}$ for a particular type of atom, can be written as  
\begin{eqnarray}
\label{MSD_01}D_{tr}&=&\lim_{t \to \infty} \frac{<d^{2}_{\alpha}>}{6t}
\end{eqnarray}
where $d_{\alpha}$ is the modulus of $\mathbf{d_{\alpha}}$ with $\mathbf{d_{\alpha}}=\mathbf{R_{j_{\alpha}}(t)}-\mathbf{R_{j_{\alpha}}(0)}$ of atom type ${\alpha}$. 
For a complicated dynamical motion, we need complex model with different contribution (such as translation, different types of rotation) to extract translational diffusion co-efficient from the spectra (experimental or simulated data). And problems seem to appear if we use a model which lacks some of the dynamical contribution (translation or rotation) but still can able to fit the data relatively well. In that case, extracted $D_{tr}$ is not correct and our conclusion about the system transport properties become wrong. One possible solution is to directly measure the CoM motion which is more straight-forward but difficult to execute in experiment. Here we show for a complex molecule like TBA$^+$, why the two above mentioned methods do not lead to same result (performed both by simulation and experiment) and how to overcome that. It is important to note that while comparing the experimental and simulation data values, our goal is not to achieve quantitative agreement but we emphasize more on qualitative comparison and the MD simulated ion translational diffusion coefficient $D_{tr}$ must be corrected by a factor of 1.24 because of the difference in viscosity effect between H$_{2}$O and D$_{2}$O~\cite{Cho99_103}. 

\begin{figure}[htbp]
\begin{center}
\includegraphics[width=0.9\textwidth,angle=0] {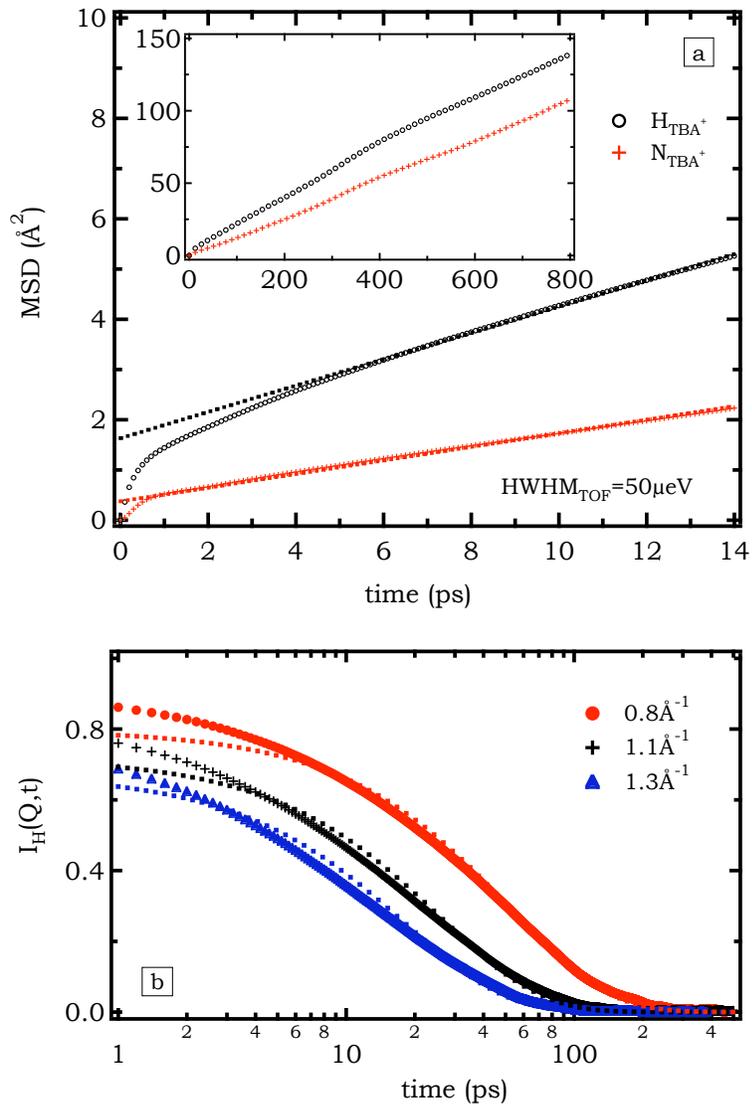}
\vspace{0.1cm}
\caption{(a) $D_{tr}$ extracted from MSD analysis of 1m (ion:water=1:56) TBABr solutions. Difference between nitrogen (CoM) motion (red cross) and average hydrogen movement (black circle) in TOF time-window (14ps). insert: MSD of the same two quantities nitrogen (CoM) and hydrogen atoms. Note the hydrogen MSD changes its slope after $\sim$350ps which indicates the global rotation time. (b) I$_{inc}(Q,t)$ of average hydrogen predicts similar D$_{tr}$ as in MSD analysis. The short time disagreement with model can arise from bond vibration, viscous relaxation, momentum relaxation or sound propagation (see section \ref{dyn_solute}.}
\label{MSD_TBA}
\end{center}
\end{figure}

\
\\
\bf{A.1 Individual Atom Motion}\normalfont
%
%
\\
\\
First we present the result of MSD. If we want to compare it with the values extracted from neutron TOF experiment, we must consider two things. First, the MSD of all the hydrogen atoms in TBA$^+$ should be calculated together. This is because the region we have exploited in our TOF experiment, is principally incoherent in nature i.e. it basically highlights average TBA$^+$ hydrogen atom dynamics (solvent is deuterated). From the plotting of TBA$^+$ hydrogen MSD (black circle) as a function of time [insert of Figure \ref{MSD_TBA} (a)], we find two clearly different slopes, one of them (which is before $\sim$400ps) shows faster motion. Considering the definition, we concentrate more at long time regime and find that $D_{tr}$ = (0.27$\pm$0.02)$\times$10$^{-9}$m$^{2}$s$^{-1}$. The same $D_{tr}$ is also tried to deduce from incoherent I(Q, t) curves [Figure \ref{MSD_TBA} (b)] with the TBA$^{+}$ hydrogen atoms using the model~\cite{Bhowmik2012}. Here we consider free continuous diffusion for overall translation of the cation with three fold jump model for terminal methyl rotation~\cite{Chahid94}~\cite{Cabral00}~\cite{Bee}. 
\begin{eqnarray}
\label{NSE_model_final}I_{inc}(Q,t)=\frac{C+B A_{0}(Qr)}{B+C}e^{-\frac{t}{\tau_{tr}}}+\frac{B[1-A_{0}(Qr)]}{B+C}e^{-t(\frac{1}{\tau_{tr}}+\frac{3}{2\tau_{rot}})}
\end{eqnarray}
where $\tau_{tr}$ and $\tau_{rot}$ are the translational relaxation and rotational time, $A_{0}(Qr)=\frac{1}{3}[1+2j_{0}(Qr)]$ with $j_{0}(Qr)$ the zeroth-order spherical Bessel function and $r$ is the H-H distance in the methyl group and $B$ and $C$ are number of hydrogen atoms in the CH$_{3}$ and CH$_{2}$ families. By these I(Q,t) data analysis we deduce $D_{tr}$ = (0.38$\pm$0.03)$\times$10$^{-9}$m$^{2}$s$^{-1}$. Obviously the two $D_{tr}$ values extracted from MSD and I(Q,t) are different. The reason is we are probing to different time regimes for MSD and I(Q,t), while for MSD is analysis is at long time (from $\sim$400ps to $\sim$800ps), the I(Q,t) curves ranges upto $\sim$200ps at most. So the second important thing is that the time window that we consider in case of our MSD analysis (in simulation) must be comparable with the energy-window of TOF experiment. Because we have mentioned before that the HWHM of our TOF experiment is 50 $\mu$eV that corresponds to $\sim$14ps ($\tau (ps)=\frac{0.673}{\Gamma (meV)}$), we only concentrate up to t=14ps in MSD data. We show that considering all the hydrogens in the ions up to 14ps, $D_{tr}$ equals to (0.40$\pm$0.03)$\times$10$^{-9}$m$^{2}$s$^{-1}$ (black dashed line of Figure \ref{MSD_TBA}). But before $\sim$5ps, we see a fast dynamics due to several effects. A rough estimate of those effects can be as follows - viscous relaxation time ($\sim$0.3ps)~\cite{Nagele96}, momentum relaxation time ($\frac {2}{9}$ of viscous relaxation)~\cite{Nagele96}, sound propagation ($\sim$0.3ps)~\cite{Nagele96}, bond vibration, methyl group rotation~\cite{Bhowmik2012}~\cite{Cabral00} etc. In Figure \ref{MSD_TBA} (a), we also have shown the estimated $D_{tr}$ derived from central nitrogen atom. We will come back to this point later and will show which one should be considered as correct cation translational diffusion coefficient. This value is comparable to MSD analysis (within a time window of t=14ps). 

In QENS experiment, we have have analyzed the individual atom motion both by the TOF and NSE experiment. From Figure \ref{S_Q} it is clear that the incoherent contribution dominates the Q region between $\sim$0.65\AA$^{-1}$ to $\sim$1.4\AA$^{-1}$. Recalling the fact that incoherent scattering cross-section for hydrogen is significantly larger than other atoms~\cite{ILL03}, we can characterize any dynamics observed in this region to the individual hydrogen atom. The incoherent TOF signal (in the quasi-elastic region) is analyzed by the same model stated earlier (equation \ref{NSE_model_final}) but in energy space $S_{inc}(Q,\omega)$~\cite{Bhowmik2012}. We find that $D_{tr }$=(0.20$\pm$ 0.03)$\times$10$^{-9}$m$^{2}$s$^{-1}$. These data are highly comparable with NMR~\cite{Ancian} and tracer diffusion~\cite{Woolf82} measurements, (0.19$\pm$ 0.01)$\times$10$^{-9}$m$^{2}$s$^{-1}$ (for tracer diffusion the difference in viscosity between H$_{2}$O and D$_{2}$O solvent is taken care of~\cite{Cho99_103}). 

\begin{figure}[htbp]
\begin{center}
\includegraphics[width=0.9\textwidth,angle=0] {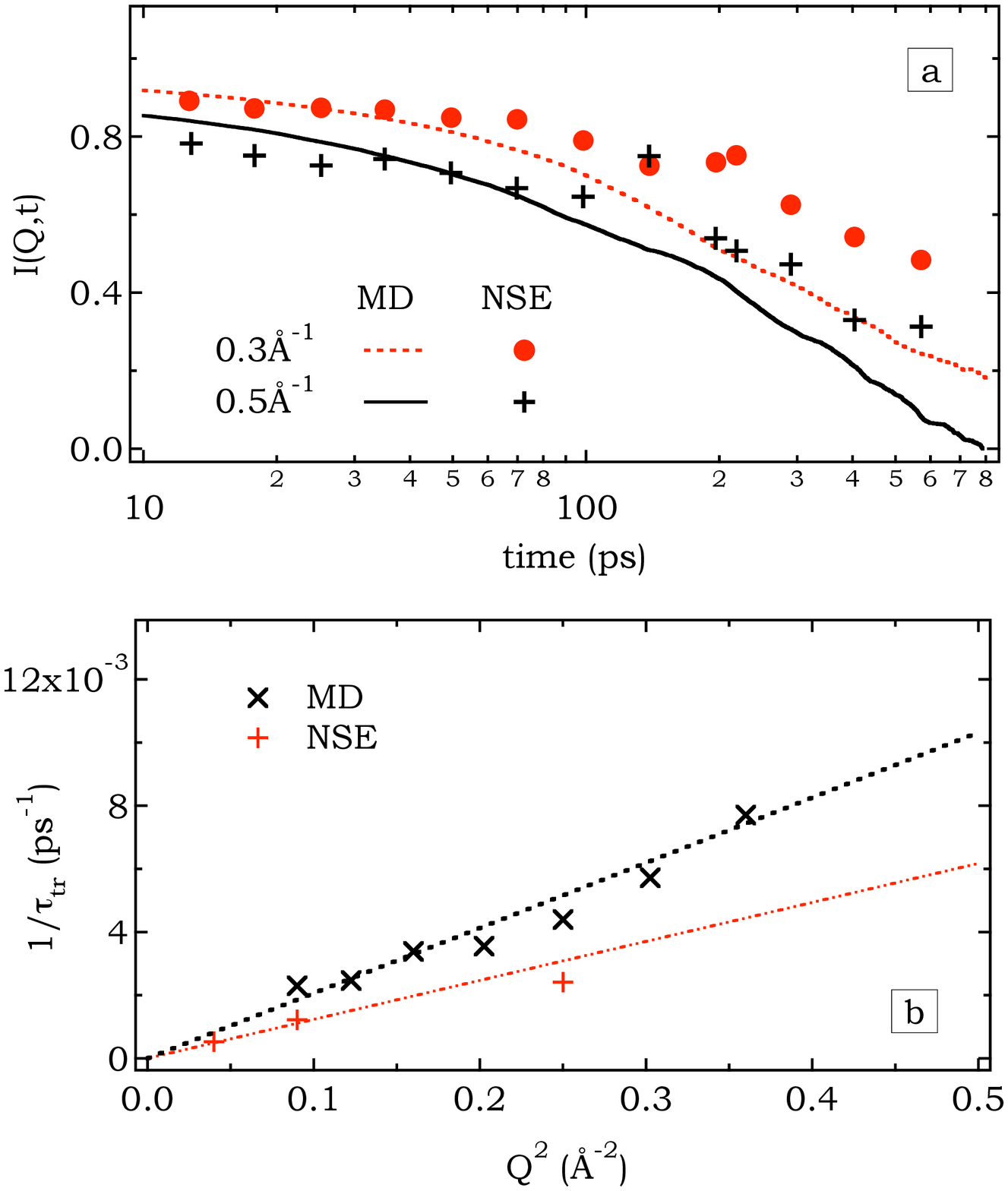}
\vspace{0.1cm}
\caption{(a) the simulated and experimentally extracted I(Q,t) curves. The simulated I(Q,t) is a combination of coherent and incoherent contribution with proportionate weightage. While comparison with NSE experiment, the simulated $D_{tr}$ should be divided by a factor of 1.24 because of the higher viscosity of heavy water (b) inverse of translational relaxation time (extracted from coherent analysis of both MD simulation and NSE experiment) is plotted against Q$^2$. $D_{tr}$ is estimated from the respective slope passing through origin. The analysis is restricted at low Q region. This allows to track the TBA$^{+}$ as a whole and ignores the rotational contribution.}
\label{I_coh}
\end{center}
\end{figure}

\
\\
\bf{A.2 Centre of Mass Motion}\normalfont
%
%
\\
\\
The second kind of motion studied is the centre of mass motion. The analysis can be done in two ways stated as before. It is clear that central nitrogen is the CoM of the TBA$^{+}$ with arms fully stretched (from our earlier SANS experiment we have calculated arms flexibility $\sim$7$\%$~\cite{Bhowmik2012}). So any diffusive motion of this nitrogen atom will denote the true CoM translational diffusive motion of the ion itself. The MSD analysis of central nitrogen predicts diffusion coefficient as $^{MSD}D^{N}_{tr}$ = (0.21$\pm$0.02)$\times$10$^{-9}$m$^{2}$s$^{-1}$ (Figure \ref{MSD_TBA}). The very short time fast motion can be due to other effects as explained before. We can also study the CoM motion by coherent intermediate scattering function [$I_{coh}$(Q,t)] as a function of time where the position correlations ($R_{j_{\alpha}}$, $R_{j_{\beta}}$) are among different and similar kinds of atoms at two distinct times (t=0 and t=t$'$)~\cite{Mezei}~\cite{Bee}~\cite{Squires}. As this is cation CoM translational motion and considering this motion as continuous diffusion, a single exponent is sufficient to extract $D_{tr}$ i.e. $I_{coh}$(Q,t) = e$^{\frac{1}{\tau_{coh}}}$ with $\tau_{coh}$ is the translational relaxation time extracted by coherent analysis. The $I_{coh}$(Q,t) curves are analyzed only at low Q (up to $\sim$0.65\AA$^{-1}$) because anything above this Q range will probe inside the ions and at very low Q rotational contribution is negligible. From $I_{coh}$(Q,t) analysis, $D_{tr}$ is estimated as (0.19$\pm$0.03)$\times$10$^{-9}$m$^{2}$s$^{-1}$[Figure \ref{I_coh} (b)].

Experimentally we have accessed the $I_{coh}$(Q,t) analysis on NSE spectrometer at low Q values ($<$0.65\AA$^{-1}$) (as the coherent contribution is predominant in this region, Figure \ref{S_Q}). Using the similar method as explained in simulation to characterize the overall cation translation, we found $D_{tr}$ is equal to (0.12$\pm$0.03)$\times$10$^{-9}$ m$^{2}$s$^{-1}$ (by a single exponential fit). In Figure \ref{I_coh} (a), we have compared the simulated and experimentally derived I(Q,t) curves at Q=0.2 and 0.3\AA$^{-1}$. These curves are the sum of incoherent and coherent contribution with their respective weightage~\cite{Mezei}~\cite{Bee}~\cite{Squires}. Note while comparing simulated and experimental I(Q,t) curves, viscosity effect must be taken care of as mentioned before in last section. Thus for simulated I(Q,t), extracted $D_{tr}$ must be divided by 1.24 (difference between water and heavy water~\cite{Cho99_103}), in other words the unit timescale of simulated I(Q,t) should increased by a factor of 1.24. 

\begin{figure}[htbp]
\begin{center}
\includegraphics[width=0.9\textwidth,angle=-90] {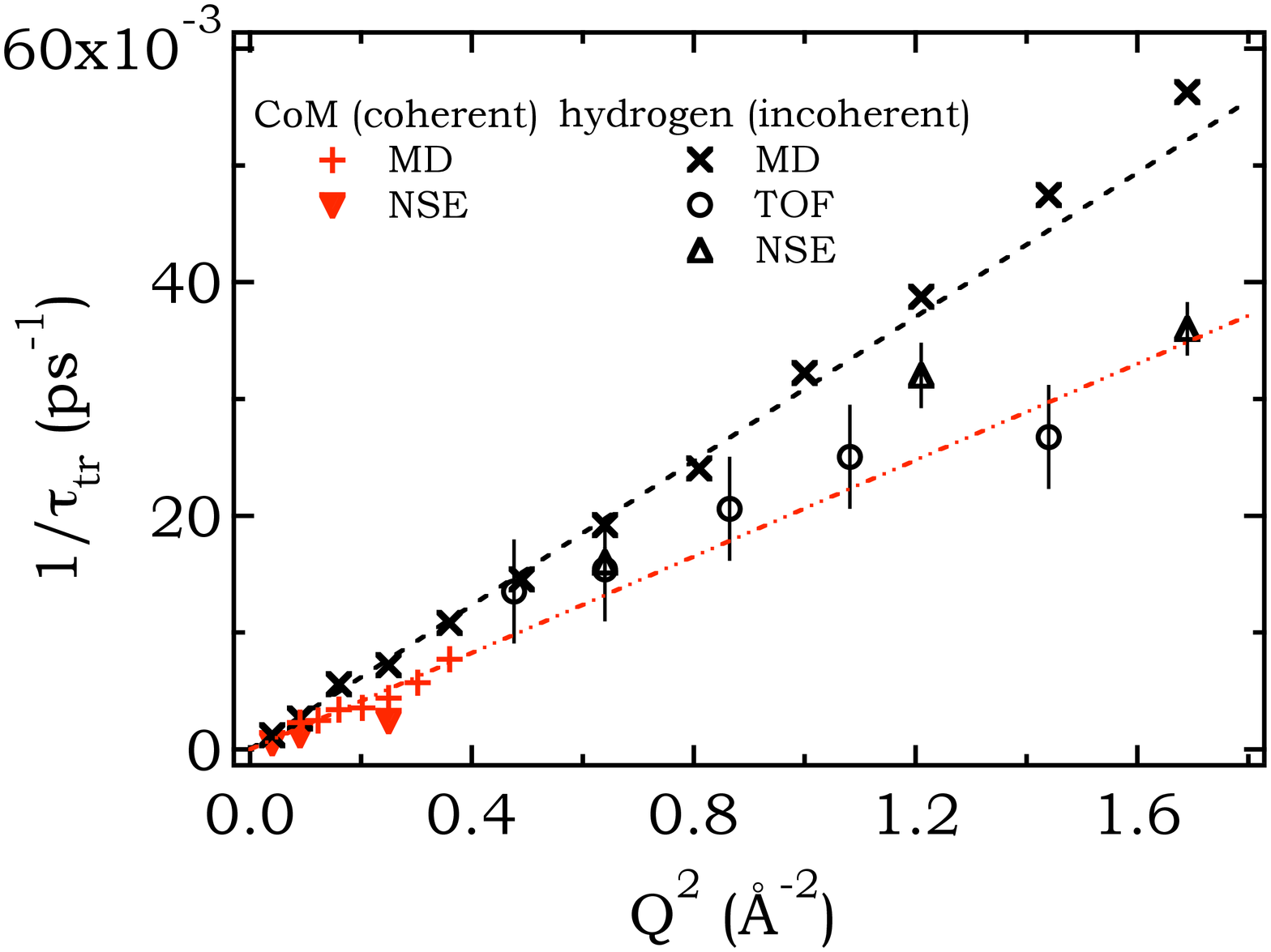}
\vspace{0.1cm}
\caption{inverse of translational relaxation time estimated both from coherent (MD and NSE) and incoherent (MD, TOF and NSE) analysis is plotted as function of Q$^2$. solute $D_{tr}$ estimated from individual hydrogen atom motion is of 2 times higher than CoM analysis (coherent treatment). Like our QENS experiment, this simulation analysis also shows the same the difference. (While comparison with experiment, the simulated $D_{tr}$ should be divided by a factor of 1.24 because of the higher viscosity of heavy water)}
\label{sim-diff}
\end{center}
\end{figure}

Thus we show the estimated $D_{tr}$ extracted from two different approaches by studying i) individual atom dynamics and ii) CoM motion. In MD simulation, the individual atom dynamics are analyzed by MSD or I(Q,t) of hydrogen atoms and experimentally the same is achieved by QENS incoherent techniques - TOF (and NSE); while CoM motion is probed by MSD (simulation) of nitrogen atoms or coherent I(Q,t) (simulation or experiment). Interestingly we see that the estimated $D_{tr}$ by coherent dynamic signal is smaller to incoherent analysis by almost a factor of 2 both in the case of simulation (Figure \ref{sim-diff}) and experiment. 

In the next segment we try to answer the discrepancy of the two different kinds of results from two different approaches. We explain it as follows - TBA$^{+}$ is itself a big cation and has four long hydrocarbon chains which also move internally. When we analyze TBA$^{+}$ hydrogen (all) atoms dynamics by a model consisting of translational and a methyl group rotation, we inevitably add the internal motion of those hydrocarbon chains into the translational part and thus the $D_{tr}$ deduced by the incoherent technique predicts a value which is higher than real translational diffusion coefficient. In coherent analysis we do not have this problem because it probes a motion where the correlations among all (same and different) atoms are counted and at very low Q any effect of rotational dynamics is negligible. With the help of coarse-grained description~\cite{Kunz92_96}~\cite{Kunz92_96_9}, it can be calculated that at low Q, most of the coherent intensity [S(Q) at less than $<$0.65\AA$^{-1}$] originates from cation-cation correlation (in the limit of Q=0, 92$\%$). Also from SANS data analysis it can be inferred that S(Q) at low Q, is mostly characterized by individual cation form factor. Thus it is a way to look at the individual TBA$^{+}$ cation as a whole to predict the overall cationic translational diffusive motion. In other words this [$I_{coh}(Q, t)$] predicts the real CoM motion. To validate our explanation we recall that MSD analysis of the central nitrogen is same to our simulated coherent $I_{coh}(Q, t)$ analysis and because the central nitrogen is the CoM of the TBA$^{+}$, any diffusive motion of this nitrogen atom will denote the real translational diffusive motion of the ion itself. Here it is important to note that all the $D_{tr}$ extracted from MD simulation should be divided by a factor of 1.24 because of the difference in viscosity between D$_{2}$O and H$_{2}$O~\cite{Cho99_103}. 

\begin{figure}[htbp]
\begin{center}
\includegraphics[width=0.8\textwidth,angle=-90] {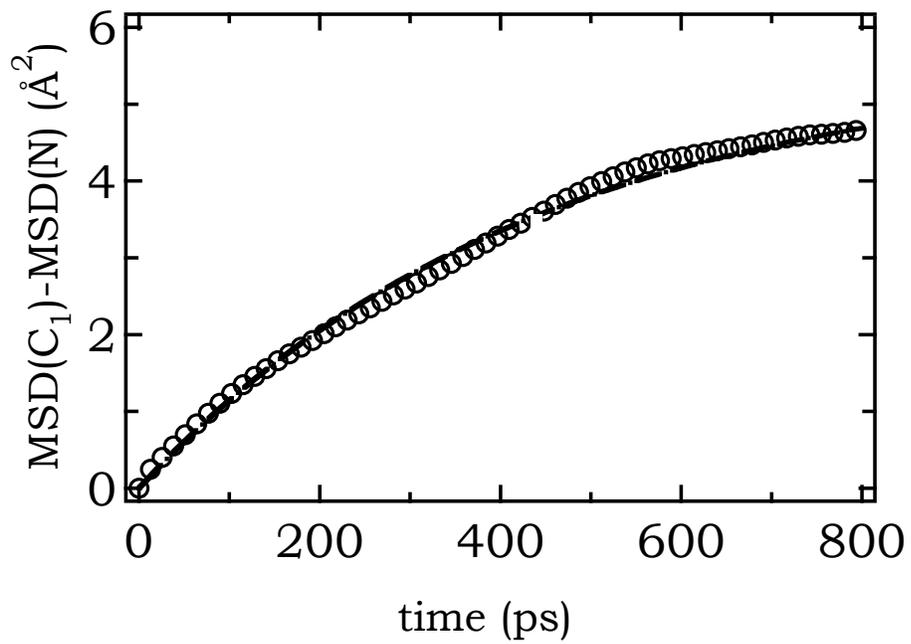}
\vspace{0.1cm}
\caption{estimation of global rotation time of TBA$^{+}$ by MSD analysis. The difference of MSD of carbon atoms (attached directly to central nitrogen) and nitrogen (CoM) is calculated which provides the global rotation time}
\label{global_rotation_TBA}
\end{center}
\end{figure}

\
\\
\bf{B. Global Motion}\normalfont
%
%
\\
\\
One added advantage of simulation is that it is easy to decouple different types of motion than experiment. As for example because of the very slow dynamics, the global rotation (of the whole cation) is almost impossible to be detected by our TOF experiment but it can be accessible by MD simulation. An estimatation of the overall rotational time, $\tau_{rot}^{glob}$ from Debye rotational time [$\tau_{rot}^{glob} = (4\pi\eta R^3)/(3k_bT)$~\cite{Paluch03}] predicts approximately 350ps (taking cation radius as $\sim$5.3\AA$ $ with D$_2$O corrected viscosity for 0.9m TBABr~\cite{Buchner02}~\cite{Eagland72}); whereas TOF resolution allows 14ps time window for measurement.
For the calculating the global rotation time we have used two different approaches. The first one is by using MSD and second one is by using the incoherent I(Q, t) curves for H atoms of TBA${+}$. 
From the structure of TBA$^{+}$, we can safely assume that the carbon atoms which are directly attached to central nitrogen, have no other motion except the translational motion and the global rotation of the cation itself. Therefore if we calculate the difference of the MSD of these carbon atoms and the central nitrogens, the translational part for carbon atoms can be eliminated and we can extract only the global rotation term. It is then fitted with $ae^{(1-\frac{t}{\tau_{rot}^{glob}})}$, where $\tau_{rot}^{glob}$ is the global rotational time and $\sqrt{\frac{a}{2}}$ is $\sim$1.5\AA$ $ as nitrogen-carbon bond length (from the fitting equation, it is easy to understand that in absence of $\tau_{rot}^{glob}$ one can have only a straight line). We estimate the $\tau_{rot}^{glob}$ $\sim$330ps (Figure \ref{global_rotation_TBA}). The effect of this global rotation can be also observed through the MSD of TBA$^{+}$ hydrogens. The change of slope can be clearly observable around 350ps (insert of Figure \ref{MSD_TBA}).
In the second method we consider the incoherent I(Q, t) curves picking only the hydrogens atoms of the cation. To extract $\tau_{rot}^{glob}$ we use a model that consists of i) translational motion of cation ($T_{tr}$) ii) terminal methyl group rotation (R$_{rot}^{met}$) iii) global rotation (R$_{rot}^{glob}$) i.e. every hydrogen experiences a translational motion; a rotational motion around the CoM and in addition to that for the terminal methyl hydrogens, there is an additional rotation around the last methyl carbon atom. Thus the complete expression can be written as
\begin{eqnarray}
\label{IQt_inc}I^H(Q,t)=\frac{T_{tr}(Q,t)}{36} [8 R_{rot,1}^{glob}(Q,t)+8R_{rot,2}^{glob}(Q,t)+8R_{rot,3}^{glob}(Q,t)+12R_{rot,4}^{glob}(Q,t)R_{rot}^{met}(Q,t)]
\end{eqnarray}
where first three terms denote the three CH$_{2}$ group hydrogen and last term denotes the methyl group hydrogen motion. We denote different types of motions as follows
\begin{eqnarray}
\label{global_fit}T_{tr}(Q, t) &=& e^{D_{tr}Q^{2}t} \nonumber \\
R_{rot,i}^{glob}(Q,t) &=& \sum^{\infty}_{l=1}(2l+1)j_{l}^{2} (Q, b)F_{rot}(t), with F_{rot}(t)=e^{-l(l+1)D_{r}t} \nonumber \\
R^{met}_{rot}(Q,t) &=& \frac{1}{3}[1+2j_{0}(Q, c)]+\frac{2}{3}[1-j_{0}(Q, c)]e^{-\frac{t}{3\tau}}
\end{eqnarray}
where $j_{l}$(Q, b) is spherical Bessel function with b refers the distance between each hydrogen and nitrogen and c is the distance between H-H distance in methyl group~\cite{Jones88}~\cite{Sears66_Hindered}~\cite{Liu02}. The fitting is done with all the 18 $I^{\alpha}_{inc}(Q, t)$ curves  at once (spaced equidistantly from 0.2\AA$^{-1}$ to 1.8\AA$^{-1}$) and calculating $l$ up to 5. This also estimates $\tau$ $\sim$330ps. It is obvious that this global rotation time is out of the TOF time window. Thus incorporating the global rotation term in TOF analysis would not change our conclusion. In case of our NSE analysis, our analysis is within the Q range between 0.8\AA to 1.4\AA where $\tau_{rot}^{glob}$ is also outside the time window. 

\subsubsection{\label{dyn_solvent}Solvent}
\begin{figure}[htbp]
\begin{center}
\includegraphics[width=0.9\textwidth,angle=0] {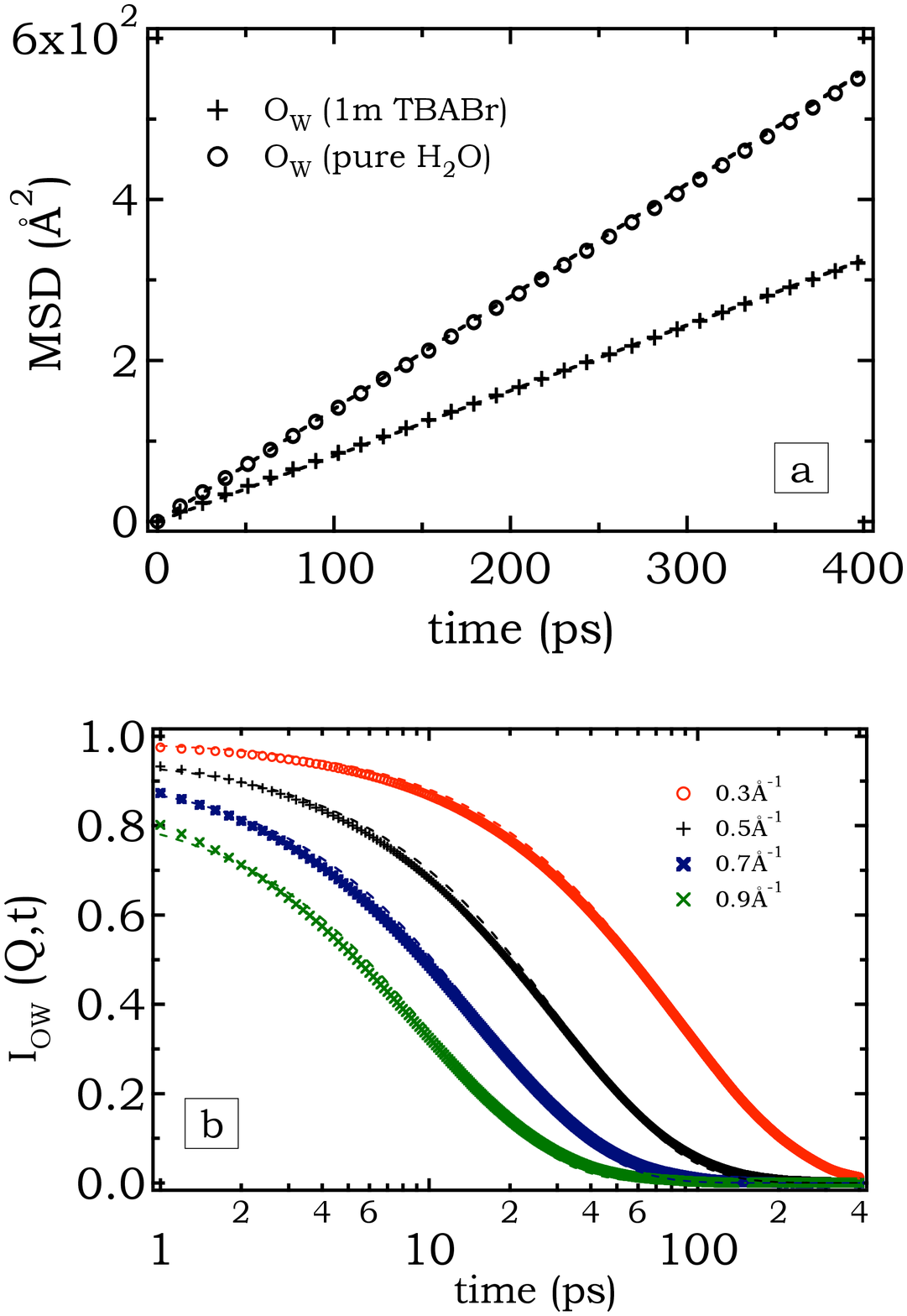}
\vspace{0.1cm}
\caption{$D_{tr}$ extracted for solvent water by (a) MSD analysis of central oxygen atom. Comparison with bulk water shows for 1m solution of TBABr, the translational diffusion is slowed by a factor of two (b) I$_{inc}(Q,t)$ of oxygen atom. It also predicts similar conclusion as MSD analysis}
\label{solvent_diff}
\end{center}
\end{figure}

\begin{figure}[htbp]
\begin{center}
\includegraphics[width=0.9\textwidth,angle=0] {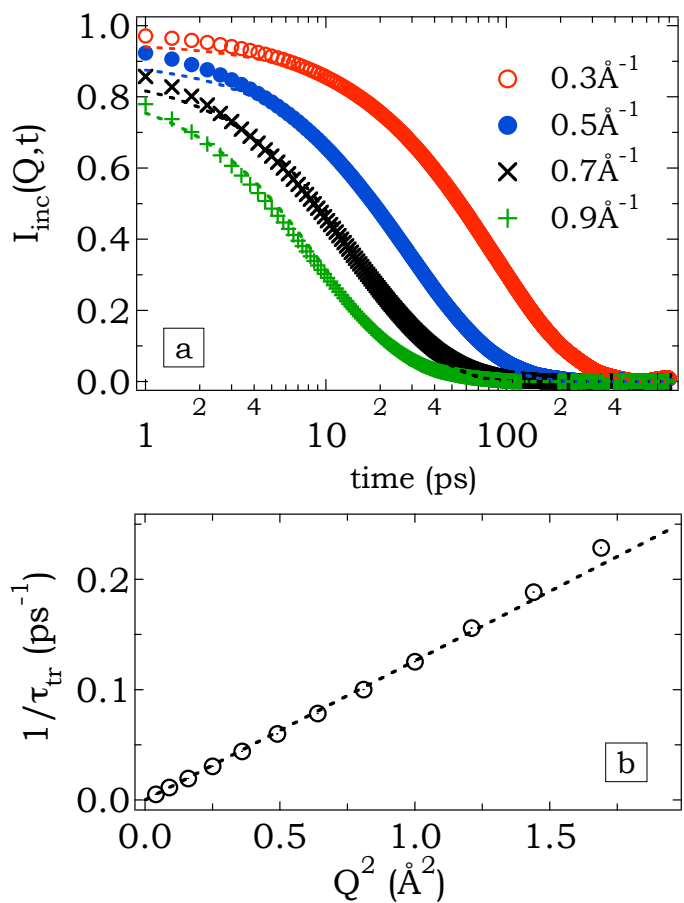}
\vspace{0.1cm}
\caption{$D_{tr}$ of solvent water from incoherent I$_{inc}(Q,t)$ analysis by MD simulation. The fitting is done with a model consisting of a translational and a rotational term, used before in QENS data treatment of bulk water~\cite{Teixeira85_31}. The result is in agreement with oxygen atom analysis}
\label{D_H2O_2}
\end{center} 
\end{figure}

%
%
Similar to solvent dynamics, we move to study the solvent dynamics or mainly its translational dynamics. If we pick the central oxygen atoms of solvent water molecules and plot their position as a function of time, the slope estimates the $D_{tr}$=(1.29$\pm$0.03) $\times$10$^{-9}$m$^{2}$s$^{-1}$ (Figure \ref{solvent_diff} (a)). The same result is obtained when the intermediate scattering function~\cite{Mezei}~\cite{Bee}~\cite{Squires} is deduced by calculating the position correlation of only the oxygen atoms (Figure \ref{solvent_diff} (b)). We find the solvent translational diffusion coefficient is $\sim$2 times smaller than that of bulk water. Now we pick only the hydrogens of water and calculate the same (Figure \ref{D_H2O_2} (a)). This time only hydrogen atoms are taken because in incoherent QENS experiment, hydrogen is more detectable than other atoms~\cite{Bee}~\cite{ILL03}. To analyze we use the model with a translational and rotational term~\cite{Teixeira85_31}. This predicts a $D_{tr}$ (1.26$\pm$0.03)$\times$10$^{-9}$m$^{2}$s$^{-1}$ with $\sim$1ps for rotational time (Figure \ref{D_H2O_2}). Note this solvent water $D_{tr}$ value is 1.5 times slower than 1m aqueous TMABr and NaBr system [(1.91$\pm$0.03)$\times$10$^{-9}$m$^{2}$s$^{-1}$], $\sim$2 times smaller than bulk water and in good agreement with earlier QENS experiment~\cite{Novokiv99_79}. 

\
\\
Table 1: Field parameters for all atom explicit TBA$^{+}$ cation
\\
\begin{tabular}{|l|ccc|}
	\hline
atom or part of molecule & atom charge & (from bold part) & \\
 & N & C & H \\
	\hline	
-\bf{N}- & 0.056669 & & \\
-N-$\bf{CH_{2}}$- & & 0.017461 & 0.053130 \\
-N-$CH_{2}-\bf{CH_{2}}$- & & -0.002556 & 0.021844 \\
-N-$CH_{2}-CH_{2}-\bf{CH_{2}}$- & & 0.011361 & 0.020886 \\
-N-$CH_{2}-CH_{2}-CH_{2}-\bf{CH_{3}}$ & & -0.086548 & 0.034799 \\
	\hline
\end{tabular}
\begin{tabular}{|c|cc|}
	\hline
bond & energy & length \\
harmonic & (Kcal/mol/\AA$^{2}$) & (\AA) \\
	\hline	
C-H$_{C}$ & 340 & 1.090 \\
C-H$_{N}$ & 240 & 1.090 \\
C-C & 310 & 1.526 \\
C-N & 367 & 1.471 \\
	\hline
\end{tabular}
\begin{tabular}{|c|cc|}
	\hline
angle & energy & angle \\
harmonic & (Kcal/mol/\AA$^{2}$) & (degree) \\
	\hline
H$_{C}$-C-H$_{C}$ & 35 & 109.5  \\
H$_{N}$-C-H$_{N}$ & 35 & 109.5\\
C-C-H$_{C}$ & 50 & 109.5 \\
C-C-H$_{C}$ & 50 & 109.5 \\
C-C-C & 40 & 109.5  \\
C-C-N & 80 & 111.2 \\
H$_{N}$-C-N & 50 & 109.5 \\
C-N-C & 50 & 109.5 \\
	\hline
\end{tabular}
\begin{tabular}{|c|cc|}
	\hline
dihedral angle & energy & angle \\
cosine & (Kcal/mol/\AA$^{2}$) & (degree) \\
	\hline
H$_{C}$-C-C-H$_{C}$ & 0.15 & 0.0 \\
H$_{C}$-C-C-C & 0.16 & 0.0 \\
H$_{C}$-C-C-H$_{N}$ & 0.15 & 0.0 \\
H$_{N}$-C-C-C & 0.16 & 180.0 \\
C-C-C-C & 0.18 & 0.0 \\
X-C-C-X & 0.15 & 0.0 \\
X-C-N-X & 0.15 & 0.0 \\
	\hline
\end{tabular}
\begin{tabular}{|c|cc|}
	\hline
atom & $\epsilon$ & $\sigma$ \\
type & Kcal & \AA \\
	\hline
H$_{C}$ & 0.0157 & 1.487  \\
H$_{N}$ & 0.0157 & 1.100\\
C & 0.1094 & 1.900 \\
N & 0.1700 & 1.8240 \\
	\hline
\end{tabular}
\\
H$_{N}$ represents the hydrogens attached to the central N

\section{\label{conclusion}Conclusion}
%
%
We study the structure and dynamics of 1m aqueous TBABr solution both by QENS technique and MD simulation at ambient temperature and compared with NaBr and TBABr. We have shown that the solvent water structure (for TBABr) is significantly different from simple salts (ex. NaBr) or other smaller TAA salts (ex. TMABr). The hydration shell is less pronounced in case of TBA$^{+}$ and the water molecules can penetrate up to the third carbon of each hydrocarbon arm (almost half the chain length). There is a non-zero probability where bromide anion can also penetrate and form ion-pair. This agrees well with earlier result and proposition~\cite{Buchner02}~\cite{Heyda10}. Our MD simulation have generated static structure factor which is comparable with other experimental techniques that we performed. By this study we have proposed a solution for the ambiguity in values of translational diffusion coefficient extracted from two different techniques (incoherent analysis estimates 2 times larger translational diffusion coefficient than coherent). We have shown that when analyzing QENS incoherent (individual H atom) data for a complex ion like TBA$^{+}$, using a simple model (combination of translation and methyl rotation) does not correctly estimate the translational diffusion coefficient. As the model fits with the data reasonably well, a more complex model is not necessary to be introduced. We have shown that this problem can be solved by coherent analysis at low Q values. By MSD analysis of central nitrogen, it is proved that coherent analysis can predict well the real CoM diffusion. The much slower global cationic rotation time is also estimated by simulation. In the last segment we discuss about the solvent dynamics and showed that in concentrated solution like one studied here, the water diffusion is reduced by a factor of two.

\section{acknowledgments}
DB specially thank to Dr. Jose T\'eixeira, Dr. Natalie Malikova, Dr. Guillaume M\'erriguet and Prof. Pierre Turq for their valuable suggestions and opinions.

\bibliography{electrolytes_24-05-2011}

\newpage


\end{document}